\documentclass[11pt,fleqn]{article}

\usepackage{cite,latexsym,amssymb,amsfonts}
\usepackage{graphicx}
\usepackage{epstopdf}

\usepackage{color}

\def\fig{Fig.\,}

\usepackage{amsfonts,amssymb,cite}

\def\noi{\noindent}

\newcommand{\Title}[1]{\noi {{\Large\bf #1}}\\[1ex]}

\def\Aunames#1{\noi{\bf #1}}
\def\auth#1{${}^{#1}$}
\def\Addresses#1{\medskip\noi \protect
	\begin{description}\itemsep -3pt {\it #1} \end{description}}
\def\addr#1#2{\item[${}^{#1}$]{\it #2}}

\newcommand{\Abstract}[1]{\vskip 2mm \begin{center}
        \parbox{16.4cm}{\small\noi #1} \end{center}\medskip}

\def\email#1#2{\footnotetext[#1]{e-mail: #2}\addtocounter{footnote}{1}}

\def\nq{\hspace*{-1em}}
\def\nqq{\hspace*{-2em}}
\def\nhq{\hspace*{-0.5em}}

\def\cm{\hspace*{1cm}}
\def\inch{\hspace*{1in}}

\def\ten#1{\mbox{$\cdot 10^{#1}$}}

\def\Jl#1#2{#1 {\bf #2},\ }

\def\ApJ#1 {\Jl{Astroph. J.}{#1}}
\def\CQG#1 {\Jl{Class. Quantum Grav.}{#1}}
\def\DAN#1 {\Jl{Dokl. AN SSSR}{#1}}
\def\GC#1 {\Jl{Grav. Cosmol.}{#1}}
\def\GRG#1 {\Jl{Gen. Rel. Grav.}{#1}}
\def\JETF#1 {\Jl{Zh. Eksp. Teor. Fiz.}{#1}}
\def\JETP#1 {\Jl{Sov. Phys. JETP}{#1}}
\def\JHEP#1 {\Jl{JHEP}{#1}}
\def\JMP#1 {\Jl{J. Math. Phys.}{#1}}
\def\NPB#1 {\Jl{Nucl. Phys. B}{#1}}
\def\NP#1 {\Jl{Nucl. Phys.}{#1}}
\def\PLA#1 {\Jl{Phys. Lett. A}{#1}}
\def\PLB#1 {\Jl{Phys. Lett. B}{#1}}
\def\PRD#1 {\Jl{Phys. Rev. D}{#1}}
\def\PRL#1 {\Jl{Phys. Rev. Lett.}{#1}}

\def\lal{&&\nqq {}}
\def\eq{Eq.\,}
\def\eqs{Eqs.\,}
\def\beq{\begin{equation}}
\def\eeq{\end{equation}}
\def\bear{\begin{eqnarray}}
\def\bearr{\begin{eqnarray} \lal}
\def\ear{\end{eqnarray}}
\def\earn{\nonumber \end{eqnarray}}

\def\nnn{\nonumber\\ \lal }
\def\nnnv{\nonumber\\[5pt] \lal }
\def\yy{\\[5pt] {}}
\def\yyy{\\[5pt] \lal }

\def\d{\partial}

\def\sign{\mathop{\rm sign}\nolimits}
\def\diag{\mathop{\rm diag}\nolimits}

\def\const{{\rm const}}

\def\then{\ \Rightarrow\ }

\def\ssph{static, spherically symmetric}

\def\elmag{electromagnetic}

\def\wh{wormhole}
\def\whs{wormholes}
\def\bu{black universe}
\def\bus{black universes}
\def\asflat{asymptotically flat}
\def\Scw{Schwarz\-schild}

\def\rf#1{(\ref{#1})}

\begin{document}
\twocolumn[

\Title{Wormholes and black universes without phantom fields\yy
           in Einstein-Cartan theory}

\Aunames{K. A. Bronnikov\auth {a,b,c,1} and A. M. Galiakhmetov\auth{d,2}}

\Addresses{ \addr a {VNIIMS, Ozyornaya ul. 46, Moscow 119361, Russia}
          \addr b {Peoples' Friendship University of Russia,
                     ul. Miklukho-Maklaya 6, Moscow 117198, Russia}
          \addr c {National Research Nuclear University ``MEPhI''
                    (Moscow Engineering Physics Institute), Moscow, Russia}
           \addr d {Donetsk National Technical University, ul. Kirova 51, 84646, Gorlovka, Ukraine}
    }

\Abstract {We obtain a family of regular static, spherically symmetric solutions in Einstein--Cartan
      theory with an electromagnetic field and a nonminimally coupled scalar field with the correct sign
      of kinetic energy density.
      At different values of its parameters, the solution, being \asflat\ at large values of
      the radial coordinate, describes  (i) twice \asflat\ symmetric \whs, (ii) asymmetric \whs\ with an
      AdS asymptotic at the ``far end'', (iii) regular black holes with an extremal horizon or two simple
      horizons, and (iv) black universes with a de Sitter asymptotic at the ``far end''. As in other
      black universe models, it is a black hole as seen by a distant observer, but beyond its horizon
      there is a nonsingular expanding universe. In all these cases, both the metric and
      the torsion are regular in the whole space.
}



\bigskip

] 
\email 1 {kb20@yandex.ru}
\email 2 {agal17@mail.ru}

\section{Introduction}

  The origin of the presently observed accelerated expansion of the Universe has become
  one of the most important problems in modern cosmology and even in theoretical
  physics as a whole. Among different theoretical models trying to explain it (see, e.g.,
  the recent reviews [1--4] and references therein), two main trends can be distinguished:
  (1) introduction of a new hypothetic form of matter with large negative pressure,
  called dark energy (DE), in the framework of general relativity (GR) (the cosmological
  constant, various kinds of quintessence, phantom matter etc.), and (2) different suggestions
  in alternative theories of gravity, such as $f(R)$ theories, multidimensional theories and
  theories involving non-Riemannian geometries, such as the Riemann-Cartan geometry
  with torsion. The simplest theory of this kind is the Einstein-Cartan theory (ECT) [5--8],
  also leading to models of accelerated expansion [9--11].

  The ECT can be considered as a degenerate version [12--14] of the Poincar\'e gauge theory
  of gravity (PGTG), in which the gravitational Lagrangian contains invariants quadratic
  in the curvature and torsion tensors. Unlike that, in the ECT the torsion is not dynamic
  since its gravitational action reduces to the curvature scalar of Riemann--Cartan
  space--time, directly generalizing the action of GR. It is nevertheless a viable theory
  of gravity: its observational predictions agree with the classical tests of GR, but it
  substantially differs from GR at very high densities of matter \cite{HO07,Trautman,BH11}.

  Theories with torsion also attract attention since torsion naturally arises in
  many approaches such as supergravity [17--19] and superstring [20--22] theories.
  One of the simplest extensions of the ECT is $f(R)$ gravity with
  torsion \cite{CCSV07,CCSV08}, and, as shown in \cite{CCSV08}, torsion can play the
  role of DE and cause an accelerated expansion of the Universe.

  Moreover,  the existence of bouncing cosmologies in the ECT \cite{G11}
  shows that torsion can replace ``exotic'' sources, violating the weak and null energy
  conditions (WEC and NEC). As is well known, such a violation is in GR a necessary
  condition for the existence of traversable wormholes \cite{hoh-vis}. Wormholes
  are a subject of particular interest as possible time machines or shortcuts between
  different universes or distant parts of the same universe, for reviews see
  \cite{vis-book, lobo-rev,  BR-book} and references therein.

  Quite a number of \wh\ solutions are known, see, e.g, [29--32] for solutions
  with minimally coupled scalar fields, \cite{br73, BVis99} for solutions with
  conformal coupling and \cite{BVis00, br-96, br-gr02} for other couplings.
  In agreement with the general results \cite{hoh-vis}, minimally coupled scalar
  fields supporting \whs\ have to be phantom (i.e., have a wrong sign of kinetic
  energy). In the case of a nonminimal coupling, there are special \wh\ solutions
  with normal fields, but in all such cases there are always regions where the
  effective gravitational constant becomes negative, that is, the
  gravitational field itself becomes a phantom \cite{br-JMP,br-star07}.
  Moreover, all such configurations, whose existence is connected with the
  phenomenon of conformal continuation \cite{br-Pol,br-JMP}, turn out to be
  unstable under radial perturbations \cite{br-gr01,br-gr02,br-gr04}.

  Some extensions of GR predict the existence of \whs\ without exotic matter,
  in particular, brane world models \cite{kb-kim1,kb-kim2}, Einstein-Gauss-Bonnet
  gravity \cite{EGB} and other high-order theories \cite{HOG-13}, the Horndeski
  theory \cite{Sush-Horn} and others.

  From the properties of the ECT it is also natural to expect that in this theory wormholes 
  can exist without exotic matter or at least without manifestly phantom fields with a 
  wrong sign of kinetic energy. 
  And indeed, a family of exact \ssph\ wormhole solutions in the ECT was 
  recently found  \cite{br-gal15} with a pair of canonical scalar fields
  as sources of gravity. One of these fields was nonminimally coupled to
  Riemann-Cartan curvature and provided the effect of torsion on the space-time metric.
  A shortcoming of these solutions was an infinite value of the torsion scalar at the
  \wh\ throat.

  Other kinds of configurations in GR whose existence is connected with NEC and
  WEC violation are regular black holes which, instead of a singularity at $r=0$, 
  contain, at the ``far end'', flat or (anti-) de Sitter asymptotic regions  \cite{BF06, bu-Dehn, 
  BBS12}. Among them of particular interest are the so-called black universes.
  By definition, a black universe is a nonsingular black hole in which, 
  beyond the event horizon, there is an expanding universe. 
  This class of models provides avoidance of singularities in
  both black holes and cosmology and combines the properties of a wormhole (absence
  of a center, a regular minimum of the area function in the case of spherical symmetry)
  and a black hole (a horizon separating static and cosmological regions of space-time).
  Moreover, the Kantowski-Sachs cosmology in the T region can be asymptotically isotropic
  and approach a de Sitter mode of expansion, which makes such models potentially
  viable for a description of an inflationary Universe or the present accelerated expansion.
  A number of such solutions of GR have been obtained with different kinds of phantom
  scalar fields as sources, with and without \elmag\ fields \cite{BF06, BBS12,
  BDon11, BKor15}.

  In the present study, we again seek \ssph\ solutions in the ECT but now with a single
  nonminimally coupled scalar field (being a source of torsion) and an \elmag\ field.
  Our purposes are (1) to obtain both \wh\ and regular black hole solutions with a
  normal scalar field, (2) to include electric or magnetic fields into consideration, and 
  (3) to avoid a singular behavior of torsion in the whole space-time.

  The paper is organized as follows. In Section 2 we present the ECT
  equations both in the general case and for static, spherically symmetric
  configurations involving an electromagnetic field and a scalar field
  nonminimally coupled to space-time curvature. Section 3 is
  devoted to finding and analyzing the properties of a family of exact
  solutions, and Section 4 is a discussion.

\section{Field equations}

  We start with the action
\bearr      \label{1}
        S = \int \sqrt{- g}d^{4}x \Bigl [- \frac{R}{2\kappa }+
        \frac{\eta }{2} \Bigl (\phi_{,k}\phi^{,k}+ \xi R\phi^2 \Bigr )
\nnn \inch
        - V(\phi ) - \frac{1}{4}F_{ik}F^{ik}\Bigr ],
\ear
  where $R = R [\Gamma]$ is the curvature scalar obtained from the full
  connection $\Gamma^{k}_{ij} = \{^{k}_{ij}\} + S_{ij\cdot }^{\ \ {k}}
  + S^k_{\cdot {ij}} + S^k_{\cdot {ji}}$; here $\{^{k}_{ij}\}$ are
  Christoffel symbols of the second kind for the metric $g_{ik}$;
  $S_{ij\cdot }^{\ \ {k}} =\Gamma^{k}_{[ij]}$ is the torsion tensor;
  $\kappa = 8\pi G$, $G$ being the Newtonian gravitational constant;
  $\phi $ is a scalar field with the potential $V(\phi)$; $F_{ik}$ is the
  Maxwell tensor. The constant $\eta = \pm 1$ corresponds to
  either a usual, canonical scalar field ($\eta = + 1$) or to a phantom one ($\eta = -1$).

  The metric $g_{ik}$ has the signature ($+\ -\ -\ -$), the Riemann
  and Ricci tensors are defined as
\[
        R^{\ \ \ m}_{ijk\cdot } = \Gamma^{m}_{jk,i} - \Gamma^{m}_{ik,j}
        + \Gamma^{m}_{ip}\Gamma^{p}_{jk} - \Gamma^{m}_{jp}\Gamma^{p}_{ik}
\]
  and $R_{jk} = R^{\ \ \ i}_{ijk\cdot }$. We should note that in ECT
  a scalar field nonminimally coupled to gravity gives rise to torsion even
  though it has zero spin. It follows from (\ref {1}) that torsion can interact with a
  scalar field only through its trace: $S_{i} = S_{ik\cdot }^{\ \ k}$
  (see \cite{KS97}). Hence, the curvature scalar $R(\Gamma) =
  g^{jk}R_{jk}$ can be presented in the form \cite{KS97}
\beq\label2
         R [\Gamma] = R[\{ \}] + 4\nabla_k S^k  -(8/3)S_k S^k  \ ,
\eeq
  where $R [\{ \}]$ is the Riemannian part of the curvature built from
  the Christoffel symbols, and $\nabla_k $ is the covariant derivative of
  Riemannian space.

  Since torsion  is induced by a nonminimally coupled scalar field only,
  the tensor $F_{ik}$ is gauge-invariant: $F_{ik} =  \d_{i}A_{k} - \d_{k}A_{i}$,
  where $A_{i}$is the potential four-vector of the electromagnetic field.

  Varying the action with the Lagrangian (\ref {1}) in $g_{ij}$,
  $S_k$, $\phi$, $\psi$ and $A_k$, we obtain the following set of equations:
\bearr             \label{3}
    G_{ij}[\{\}] = \kappa (T_{ij}[\phi] + T_{ij}[e]) + \Lambda_{ij},
\yyy         \label{4}
    S^k = \frac{3}2  \xi \Psi \phi \phi^{,k},
\yyy     \label{5}
    \Box \phi - \xi \phi R [\Gamma] + \eta dV/d\phi = 0,
\yyy         \label{6}
    \frac{1}{\sqrt{- g}}\d_i  (\sqrt{- g}F^{ik}) = 0,
\ear
  where
\bearr         \label{7}
    T_{ij}[\phi] = \eta \biggl \{\phi_{,i}\phi_{,j}
                 - \frac 12 \Big[\phi_{,m}\phi^{,m} + \xi R[\{ \}]\phi^2
\nnn \ \ \
          - 2\eta V(\phi )\Big]\, g_{ij}
                 + \xi \Big[-4S_{(i}\nabla_{j)} + 2g_{ij}S^{n}\nabla_{n}
\nnn \ \ \
          - \nabla_{i}\nabla_{j} + g_{ij}\Box + R_{ij}[\{ \}] -
               \Lambda_{ij}\Big ]\phi^2 \biggr\},
\yyy               \label{8}
    T_{ij}[e] = - F^{k}_{i}F_{jk} + \frac{1}{4}F_{lk}F^{lk}g_{ij},
\yyy             \label{9}
    \Lambda_{ij} = \frac{8}{3} S_{i}S_{j} - \frac{4}{3} S_k S^k g_{ij}.
\ear
  Here $\Box$ is the d'Alembertian operator of Riemannian space, and we
  denote $\Psi =\kappa (\eta - \kappa \xi \phi^2 )^{- 1}$.

  It is easy to verify that the effective scalar-torsion
  stress-energy tensor (SET) $T^{\rm (eff)}_{ij}[\phi]$
\beq\label{10}
        T^{\rm (eff)}_{ij} [\phi] = T_{ij}[\phi] + \kappa^{- 1}\Lambda_{ij}
\eeq
  as well as the electromagnetic SET $T_{ij}[e]$  are
  separately covariantly conserved since no explicit coupling is assumed
  between the scalar and electromagnetic fields:
\beq\label{11}
         \nabla^j T^{\rm (eff)}_{ij} [\phi] = \nabla^j T_{ij}[e] = 0.
\eeq

  The general static, spherically symmetric metric can be written in
  the form
\beq\label{12}
           ds^2  = A(u)dt^2  - \frac{du^2}{A(u)} - r^2 (u)d\Omega^2
\eeq
  in terms of the so-called quasiglobal radial coordinate \cite{BR-book},
  where $g_{00} = A(u)$ may be called the redshift function while $r(u)$ is
  the area function, or spherical radius; $d\Omega^2  = d\theta^2  + \sin^2 \theta d\varphi^2$
  is the linear element on a unit sphere. (As usual, the metric is only
  formally static: it is really static if $A > 0$, but it describes a
  Kantowski--Sachs (KS) type cosmology if $A < 0$, and $u$ is then a temporal
  coordinate.) We also consider $\phi = \phi(u)$.

  The nonvanishing components of the effective scalar-torsion SET are given by
\bearr             \label{13}
    T^{1\ \rm (eff)}_1 [\phi] = \eta \xi \phi^2 G^1_1 [\{\}] + Y + \eta \Big[ 2\xi \phi \phi''A
\nnn \quad \
    + \xi \phi \phi'A'+ (- 1 + 2\xi + 6\xi^2 \phi^2 \Psi)\phi'^2 A \Big],
\yyy                \label{14}
    T^{2\ \rm (eff)}_2 [\phi] = T^{3 \rm (eff)}_3 [\phi]
\nnn \quad \
    = \eta \xi \phi^2 G^2_2 [\{\}] + Y + 2\eta \xi \frac{r'}{r}\phi \phi'A,
\yyy                      \label{15}
    T^{0 \rm (eff)}_0 [\phi] = \eta \xi \phi^2 G^0_0[\{\}] + Y + \eta \xi \phi \phi'A',
\ear
  where the prime means $d/du$ and
\bearr             \label{16}
    Y = V(\phi ) + \eta A \biggl[ -2\xi \phi \phi''  -
    2\xi \Bigl (\frac{A'}{A} + \frac{2r'}{r}\Bigr )\phi \phi'
\nnn \inch
    + \Bigl (\frac 12  - 2\xi - 3\xi^2 \phi^2 \Psi \Bigr)\phi'^2 \biggr],
\ear
  and $G^i_k [\{\}]$ is the Einstein tensor of Riemannian space.

  The electromagnetic field compatible with the metric (\ref {12}) can
  have the nonzero components
\[ \nq
   F_{01} = {-} F_{10} \ \mbox{(electric)\ and}\  F_{23} = {-} F_{32} \ \mbox{(magnetic)},
\]
  such that
\beq                      \label{17}
      F_{01}F^{01} = - \frac{q^2_e}{r^4 (u)}, \quad
      F_{23}F^{23} = \frac{q^2_m}{r^4 (u)},
\eeq
  where the constants $q_{e}$ and $q_{m}$ are the electric
  and magnetic charges, respectively. The corresponding SET is
\beq\label{18}
     T^k_i [e] = \frac{q^2 }{r^{4}(u)}\diag(1, 1, - 1, - 1) \ ,
\eeq
  where $q^2  = q^2 _{e} + q^2 _{m}$. Thus the electromagnetic
  field equations have already been solved.

  The scalar field equation and three independent combinations of the
  Einstein--Cartan equations read
\bearr                            \label{19}
    (1 - 6\xi^2 \phi^2 \Psi )\frac{(Ar^2 \phi')'}{r^2 } -
    \frac{6\eta }{\kappa }\xi^2 \phi\phi'^2 \Psi^2 A
\nnn \ \ \
    - \eta \frac{dV}{d\phi }
    + \xi \phi \biggl [ A'' + \frac{4Ar''}{r} +\frac{4A'r'}{r}
\nnn \inch
    + \frac{2Ar'^2 }{r^2 } - \frac2 {r^2 }\biggr] = 0,
\yyy
    (A'r^2 )' = - 2\eta r^2 V\Psi +
    \frac{\eta q^2 }{r^2 }\Psi
\nnn \ \ \      \label{20}
    + 2\xi r^2 A\Psi \biggl[\phi \phi'' + \phi'^2  +
    2\Bigl (\frac{A'}{A} + \frac{r'}{r}\Bigr )\phi\phi' \biggr],
\yyy                       \label{21}
    2\frac{r''}{r} = \Psi \Big[ 2\xi \phi \phi''
    + (- 1 + 2\xi + 6\xi^2 \phi^2 \Psi )\phi'^2 \Big],
\yyy        \label{22}
    A(r^2)'' - r^2 A'' = 2 - \frac{2\eta q^2 }{r^2 }\Psi
\nnn \inch
    + 2\xi r^2 \phi\phi'\Psi \Bigl(\frac{2Ar'}{r} - A' \Bigr).
\ear
  It should be noted that the scalar field equation (\ref{19}) follows from (\ref{20})--(\ref{22}).

  We see that in the Einstein-Cartan equations (\ref{19})--(\ref{22}) the
  terms induced by torsion contain the factor $\xi^2$, i.e., they exist due
  to nonminimal coupling of the $\phi$ field with space-time curvature.

\section{Exact solution}

  Let us now try to  solve \eqs (\ref{20})--(\ref{22}).
  Assuming $\eta = + 1 $ and $\xi > 0$, the form of the expression
  for $\Psi $ and \eq (\ref{21}) prompts us to choose the following ansatz for $\phi $:\footnote
            {In a similar way one can obtain solutions for $\eta = -1$ corresponding to a
              phantom scalar, but they are beyond the scope of this paper.}
\beq   \label{23}
   \phi (u) = \frac{1}{\sqrt{\kappa \xi }}\frac{u}{\sqrt{u^2  + b^2 }} =
   \frac{1}{\sqrt{\kappa \xi }}\frac{x}{\sqrt{x^2  + 1}} \ ,
\eeq
  where $x = u/b$, and $b > 0$ is an arbitrary constant (the
  length scale). As a result, \eq (\ref{21}) takes the form
\beq\label{24}
           \frac{r''}{r} = \frac{2\xi - 1}{2\xi (x^2  + 1)^2 } \ ,
\eeq
  where the prime now stands for $d/dx$. Since in \wh\ and \bu\ solutions the radius $r(x)$
  should grow to infinity at both ends of the $x$ range, only solutions with $r'' > 0$ are
  admissible, therefore we choose $\xi > 1/2$.

  The general solution of \eq (\ref{24}) is \cite{Kamke}
\bearr           \label{25}
          r(x) = b\sqrt{x^2 + 1}\ [\gamma \cos z + \beta \sin z]
\nnn \cm
          \equiv \gamma b\sqrt{x^2 + 1}\cos z,
\nnnv
               z := p\arctan x + \alpha, \quad \ p := (2\xi )^{- 1/2} < 1,
\ear
  where $\alpha$, $\beta$, $\gamma$ are integration constants.
  To obtain $r > 0$ at all $x$, we suppose $\gamma > 0$, and since there is an arbitrary
  factor $b$, without loss of generality we put  $\gamma =1$. We also put
\beq \nhq               \label{25a}
            -\frac{\pi}{2}(1 {-}p) \leq \alpha \leq \frac{\pi}{2} (1 {-} p)
      \  \then \  p \leq 1 - \frac {2 |\alpha|}{\pi}
\eeq
  (the second inequality is useful if $\alpha$ is known). Under these conditions 
  (with strict inequality) $\cos z >0$ at all finite $x$, which leads to globally regular
  solutions.

\begin{figure}
\centering
\includegraphics[width=7.5cm,height=5.5cm]{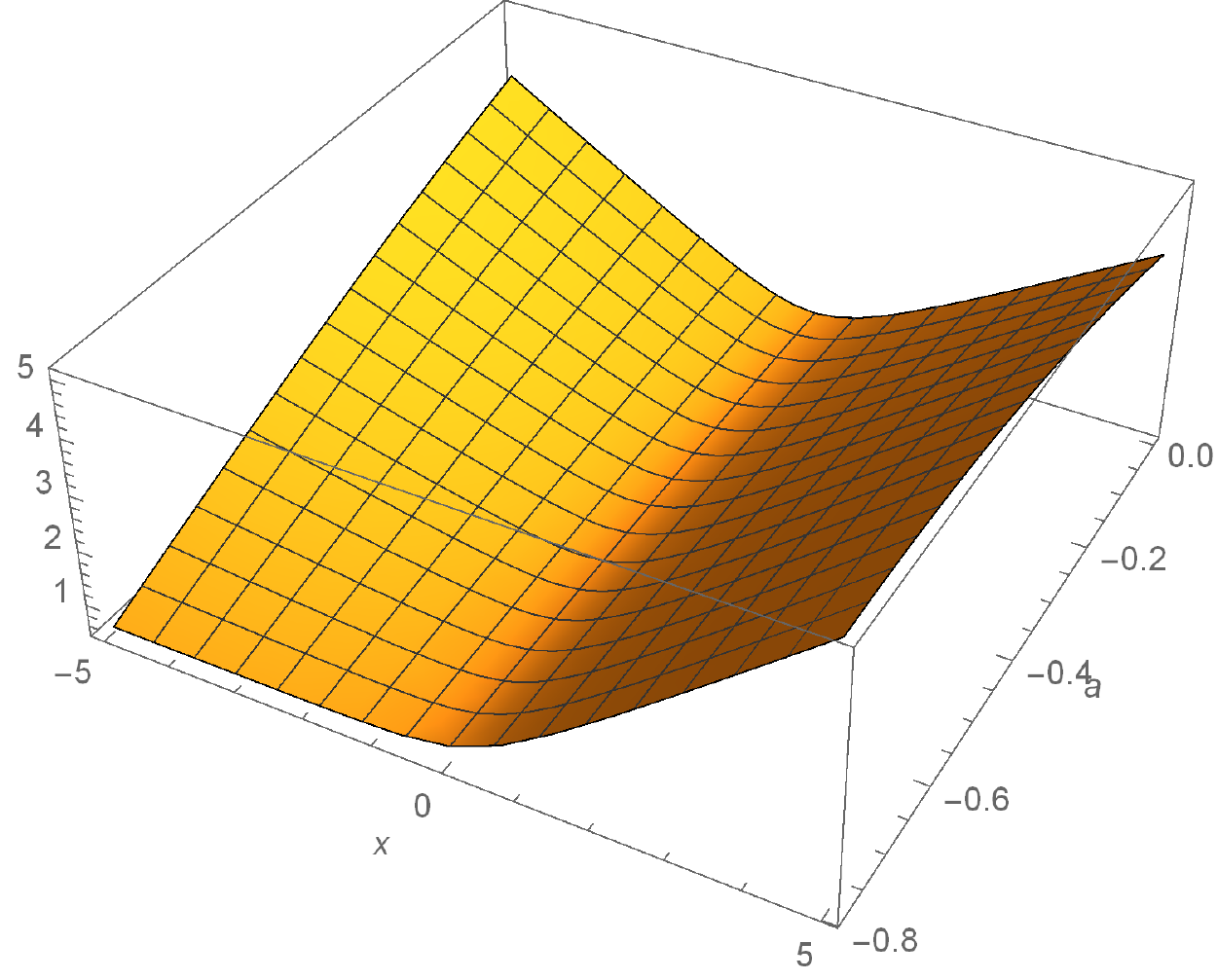}
\caption{\small The function $r(x)$ for $p=1/2$ and $- \pi/4 \leq
\alpha \leq 0$. }\label{r(x)}
\end{figure}

  From (\ref{25}) it follows
\beq         \label{26}
      \frac{r'}{r} =  \frac{x - p\tan z}{x^2  + 1} ,
\eeq
  so that  $r' = 0$ where $x = p \tan z$. This value of $x$ is a minimum of $r(x)$
  inside the range \rf{25a} of $\alpha$, while at its ends it is a limiting value at
  one of the infinities, see \fig\ref{r(x)}. This corresponds to what is sometimes called
  a ``horn'': at one end $r$ tends to a constant.
  The function $r(x)$ is even if $\alpha =0$.

  The other metric function $A(x)$ is found from \eq (\ref{22}) which can be
  presented as
\bearr \label{27}
       B'' + 2B'\ \frac{x - 2p\tan z}{x^2  + 1}
                           + \frac{2}{(x^2  + 1)^2 \cos^4 z}
\nnn \cm
                          -  \frac{2 Q}{(x^2  + 1)^2 \cos^6 z}  = 0 ,
\ear
  where $Q = \kappa q^2 /b^2$ and
\beq    \label{28}
    B(x) = \frac{A(x)}{(x^2  + 1)\cos^2 z}.
\eeq
  Its integration gives
\bearr                               \label{29}
    B(x) = B_0  + \frac{1}{3 p^2 }\Bigl[\sec^2 z - 4\ln \cos z + \frac 32 Q \sec^4 z
\nnn \cm
                -2 (K+z)  (2 + \sec^2 z)\tan z\Bigr ],
\ear
  where $B_0 $ and $K$ are integration constants.

  It is easy to see that in the general case (except for the limiting cases $\alpha = \pm (1-p)\pi/2$)
  $B(x)$ tends to finite limits as $x\to \pm\infty$,
  which, according to \rf{28}, leads to $A \sim x^2$ at both infinities, i.e., to de Sitter (dS)
  or AdS asymptotic behaviors. We will, however, restrict ourselves to discussing only systems
  which are asymptotically flat systems as $x\to + \infty $. Hence $B(x) \sim x^{-2}$, and
  in the expansion of $B(x)$  in inverse powers of $x$ the first two terms, $O(1)$ and
  $O(1/x)$, should vanish, which fixes the constants $B_0$ and $K$:
\bearr                                 \label{30}
       6 p^2 B_0  =  8\ln \cos a -  2\sec^2 a
\nnn \quad\
     + Q\Big [{-} 3 \sec^4 a + 4 \tan^2 a(2 + \sec^2 a)\Big],
\yyy                    \label{31}
       K = -a + Q \tan a
\ear
  where
\beq                    \label{32}
            a = \alpha + \pi p/2.
\eeq
  Under the conditions \rf{30} and \rf{31}, we have the following expression for $A(x)$
  at large positive $x$:
\bearr              \label{33}
        A(x) = \Big(1+ \frac{2p}{x}\tan a\Big)\biggl[\frac{Q - \cos^2 a}{\cos^4 a}
\nnn\cm
        -\frac{2p (5Q - 4\cos^2 a)}{3x \cos^5 a}\sin a\biggr] + O(x^{-2}).
\ear
  It follows from (\ref{33}) that the system can be asymptotically flat only if $Q > \cos^2 a$,
  that is, only in the presence of an electromagnetic field.

  Choosing properly the time scale at infinity and taking into account that
  $r \approx  b x\cos a$ at large $x$, we obtain from \rf{33} a
  Schwarzschild-like form of $g_{00}$:
\beq                    \label{34}
            \frac {A(x)}{A(\infty)} = 1 - \frac{2pb}{3r} \frac{2Q - \cos^2 a}{Q - \cos^2 a}\sin a,
\eeq
  and comparing it with the expression $1- 2Gm/r$, we see that the Schwarzschild mass is
\beq                \label{35}
            m = \frac{bp}{3 l_{\rm pl}}m_{\rm pl}\frac{2Q - \cos^2 a}{Q - \cos^2 a}\sin a.
\eeq
  where $m_{\rm pl} = 1/\sqrt{G}$ and $l_{\rm pl} = \sqrt{G}$ are the
  Planck mass and length, respectively. If we require $m \geq 0$, then
  from (\ref{35}) it follows $a \geq 0$.

  From \eq (\ref{20}) one derives the potential $V(x)$:
\bearr     \label{36}
        V(x) = \frac{1}{3\kappa b^2 (x^2 + 1)^2}\Bigl [4
\nnn \cm
           + (3 p^2 B_0 - 4\ln \cos z)(4\cos^2 z - 3)
\nnn \cm
           - 4(K + z) (4\cos^2 z - 1)\tan z\Bigr ] .
\ear
  To express $V(x)$ in terms of $\phi $, by \rf{23}, we must put
  $x = \sqrt{\kappa \xi }\phi (1 - \kappa \xi \phi^2 )^{-1/2}$.

  Lastly, the expression for the squared trace of torsion $S^2  = S_{k}S^{k}$
  has the form
\beq                          \label{37}
                     S^2  = - \frac{9x^2 A(x)}{4b^2 (x^2  + 1)^2 } \ .
\eeq
  Thus $S^i$ is a spacelike vector for $A > 0$ and a timelike one
  for $A < 0$, i.e., in a KS cosmology. For the solution  \rf{25}, \rf{29}
  the torsion is everywhere regular and finite; it is zero at $x=0$ which is a
  minimum of $r(x)$ if $\alpha = 0$. At flat infinity the torsion invariant \rf{37} decays
  by the law
\beq   \label{38}
            S^2 \Big|_{x\to + \infty }\sim x^{- 2} \to 0 \ .
\eeq
  At a dS/AdS infinity the invariant \rf{37} tends to a nonzero constant.

  Thus our solution  \rf{25}, \rf{29} under the conditions \rf{30}, \rf{31}
  contains two integration constants $\alpha$ (or $a$) and $K$, which determine the mass
  $m$ and the dimensionless charge $Q = \kappa q^2/b^2$. There are also two free parameters
  $b > 0$ (the length scale) and $p = (2\xi)^{-1/2} \in (0,1)$ characterizing the nonminimal
  coupling of the scalar field to curvature.

  Of interest is the asymptotic value $B(-\infty)$: if it is negative, then the whole
  configuration is a black universe, otherwise it is either a wormhole (if $B(x) >0$ everywhere)
  or a regular black hole with a static region at large negative $x$.
  The expression for $B(-\infty)$ is very simple if we put $\alpha=0$,
  in which case the function $r(x)$ is even while $z(x)$ is odd.
  Indeed, under the conditions \rf{30}, \rf{31} we have
\beq                      \label{B-}
            B(-\infty) = \frac{4K}{3p^2} \tan a (2+ \sec^2 a)
\eeq
  By \rf{35}, $\sign m = \sign a$, hence if $m >0$, the sign of $B(-\infty)$ is determined
  by the sign of $K = Q\tan a - a$. For $Q$ we only have the inequality $Q > \cos^2 a$, which
  allows the expression $Q\tan a - a$ to have any sign. We conclude that {\it our solution can
  describe black universes with $m \geq 0$}.
\begin{figure*}
\centering
    \includegraphics[width=6.5cm,height=4.8cm]{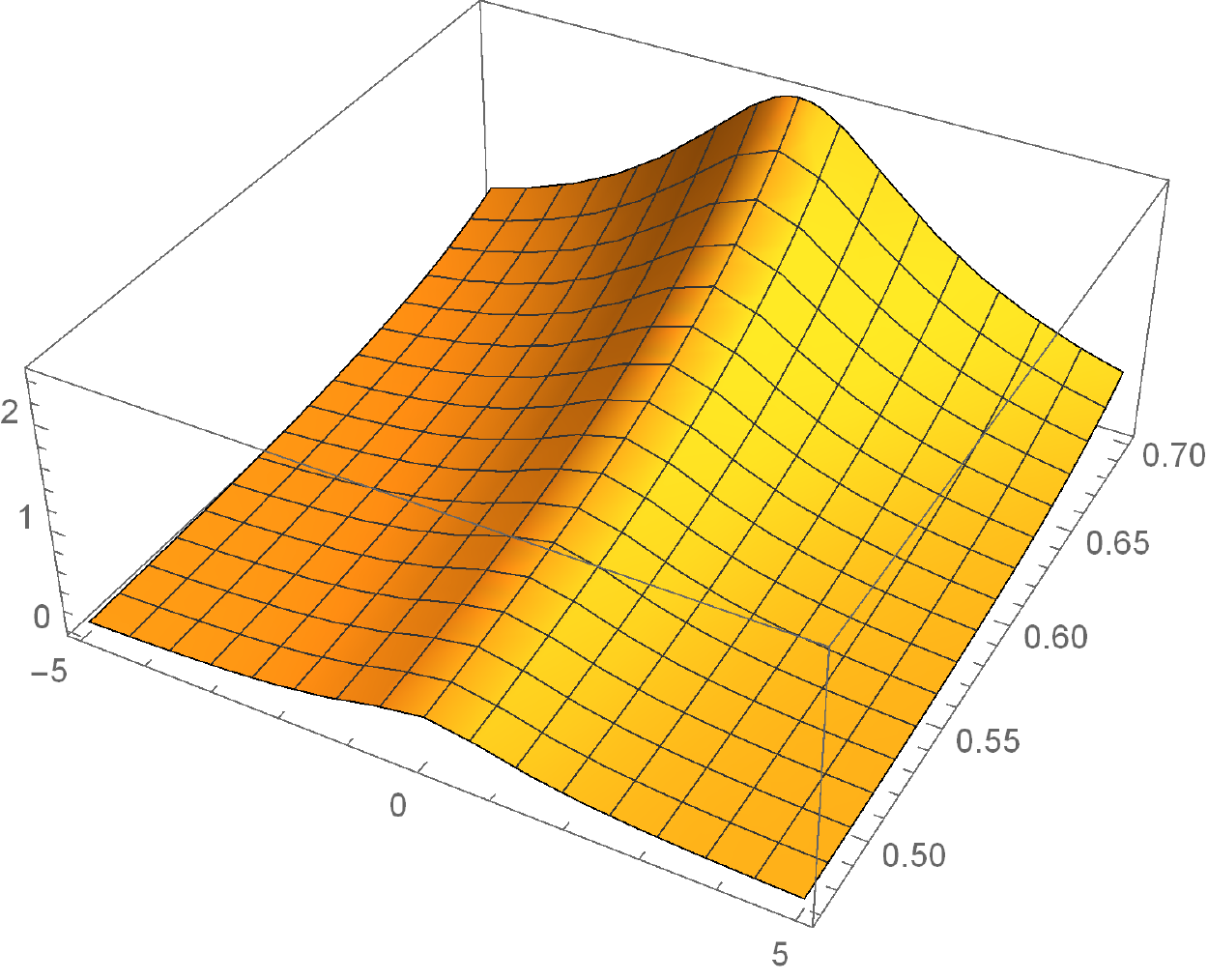}
\qquad
    \includegraphics[width=6.5cm,height=4.8cm]{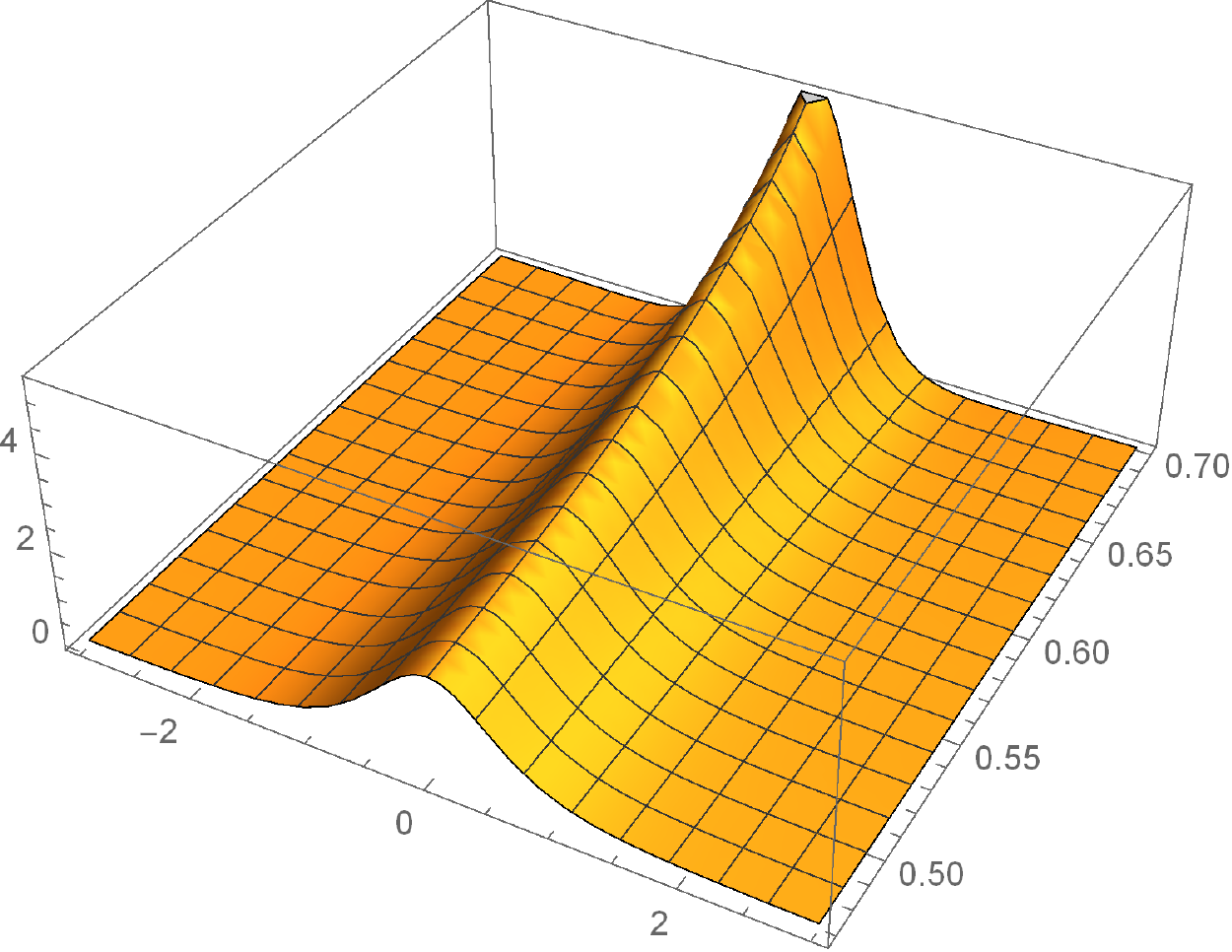}
\caption{\small
   Plots of $B(x)$ (left) and $V(x)$ (right) for symmetric configurations
    with $p$ ranging from 0.47 to 0.7.}\label{f2}
\bigskip
\end{figure*}
\begin{figure*}
\centering
\includegraphics[width=6.5cm,height=4.8cm]{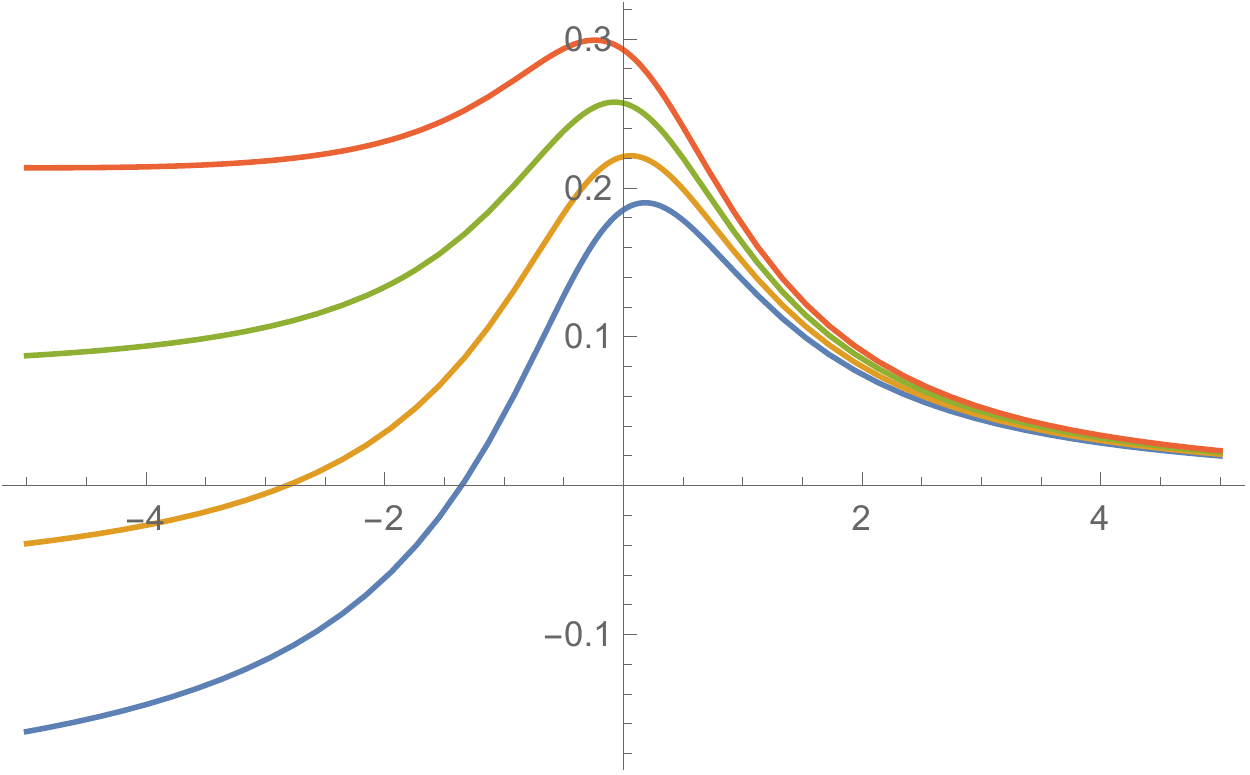}
\qquad
\includegraphics[width=6.5cm,height=4.8cm]{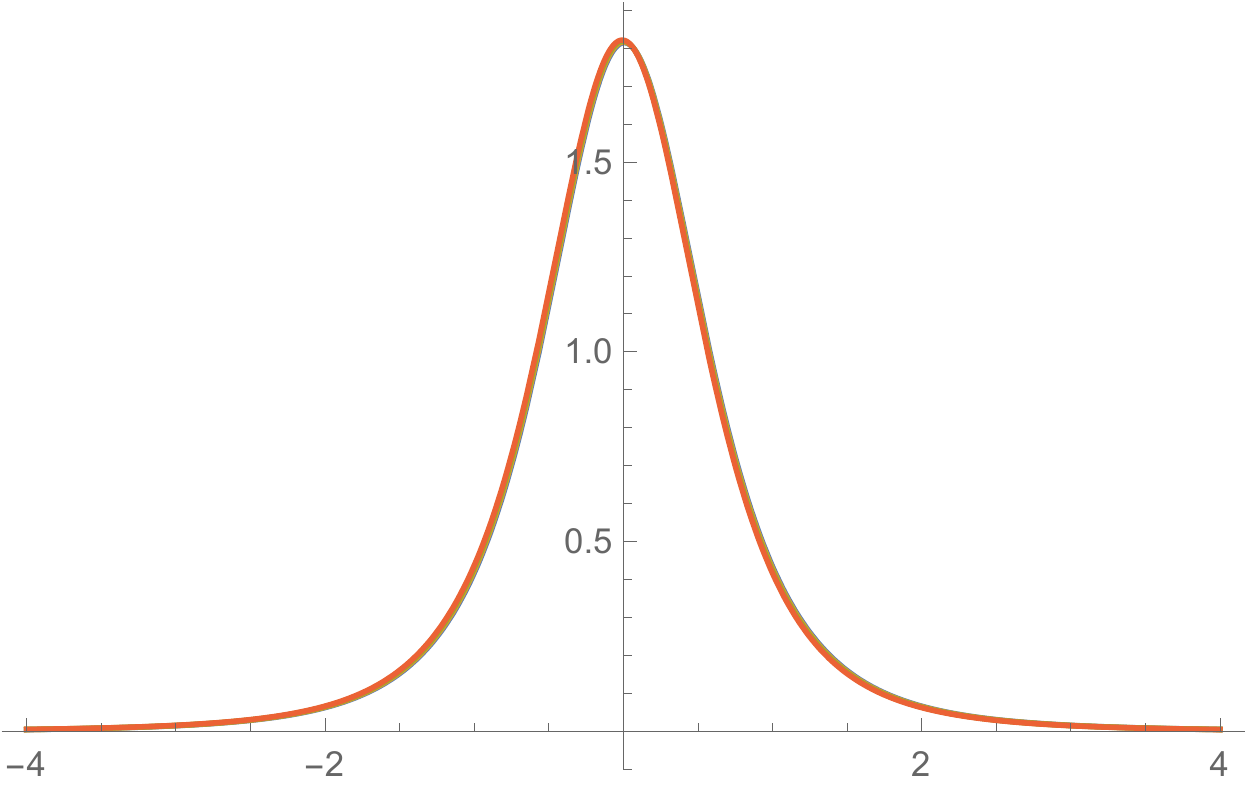}
\caption{\small
    Plots of $B(x)$ (left) and $V(x)$ (right) for asymmetric configurations with even $r(x)$,
    and the parameters $p = 0.4$, $Q = 0.85,\ 0.86,\ 0.87,\ 0.88$ (bottom-up). The plots of
    $V(x)$ almost merge for these close values of the parameters though the corresponding
    geometries are qualitatively different.}\label{f3}
\bigskip
\end{figure*}

  In what follows we will briefly describe the properties of the solution for different
  values of its parameters under the conditions \rf{30}, \rf{31}.

\subsection{Symmetric configurations}

  The solution is symmetric under the reflection $x \mapsto -x$ if and only if $\alpha =0$
  and $K=0$. Then the only free parameters are the length scale $b$  and $p$, so that
\bearr                 \label{sym}
      a = \pi p/2, \quad\  Q = a \cot a,
\nnn
           Gm = \frac {2ab}{3\pi} \sin a \frac{4a - \sin (2a)}{2a - \sin (2a)}.
\ear
  The solution exists for all $a \in (0, \pi/2)$, or $p \in (0, 1)$, or $\xi  > 1/2$, and
  $Gm \in (b/\pi, 2b/3)$.

  Plots of $B(x)$ and\footnote
    {Here and henceforth we actually plot the function $3\kappa b^2 V(x)$
           instead of $V(x)$.}
  $V(\phi)$ for different $p$ are shown in \fig{\ref{f2}}.
  Such symmetric configurations are asymptotically flat at both ends,
  $x\to \pm \infty $, and represent twice asymptotically flat (symbolically, M--M
  where ``M'' stands for ``Minkowski'') traversable wormholes.
  Given the value of $b$, which is the throat radius, the Schwarzschild mass at both ends,
  found according to (41), is the smallest at $a \ll 1$, at which
\[
    Gm \approx Gm_{\min} = b/\pi.
\]
  It is easy to estimate that if we suppose that the wormhole is large enough for
  transportation purposes, say, $b=10$ m, then this minimum mass will be about
  $2.62 \ten{30}$ g, about 440 Earth's masses. The gravitational field in such a wormhole
  will be evidently too strong for a human being to survive.

\begin{figure*}
\centering
\includegraphics[width=6.5cm,height=4.8cm]{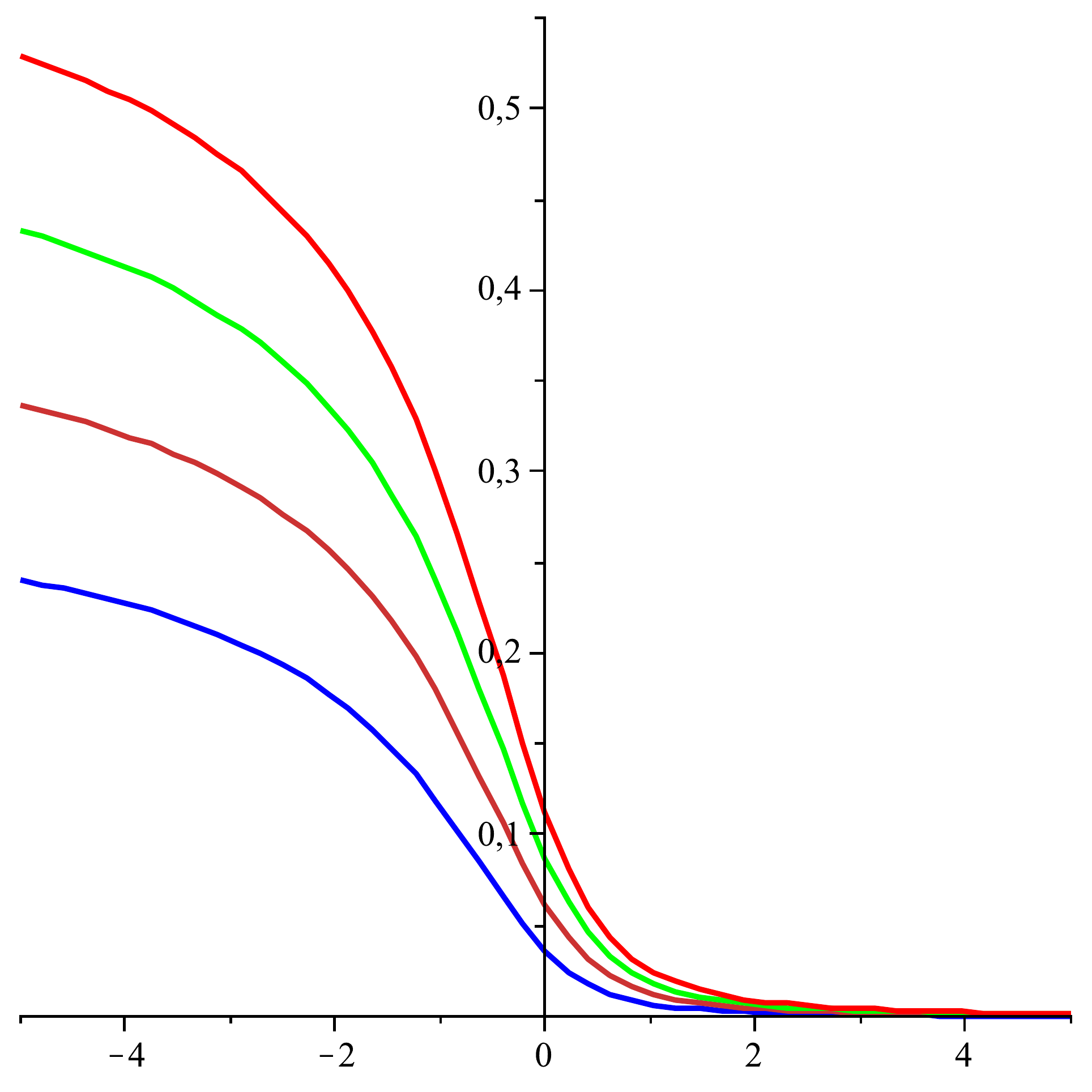}
\qquad
\includegraphics[width=6.5cm,height=4.8cm]{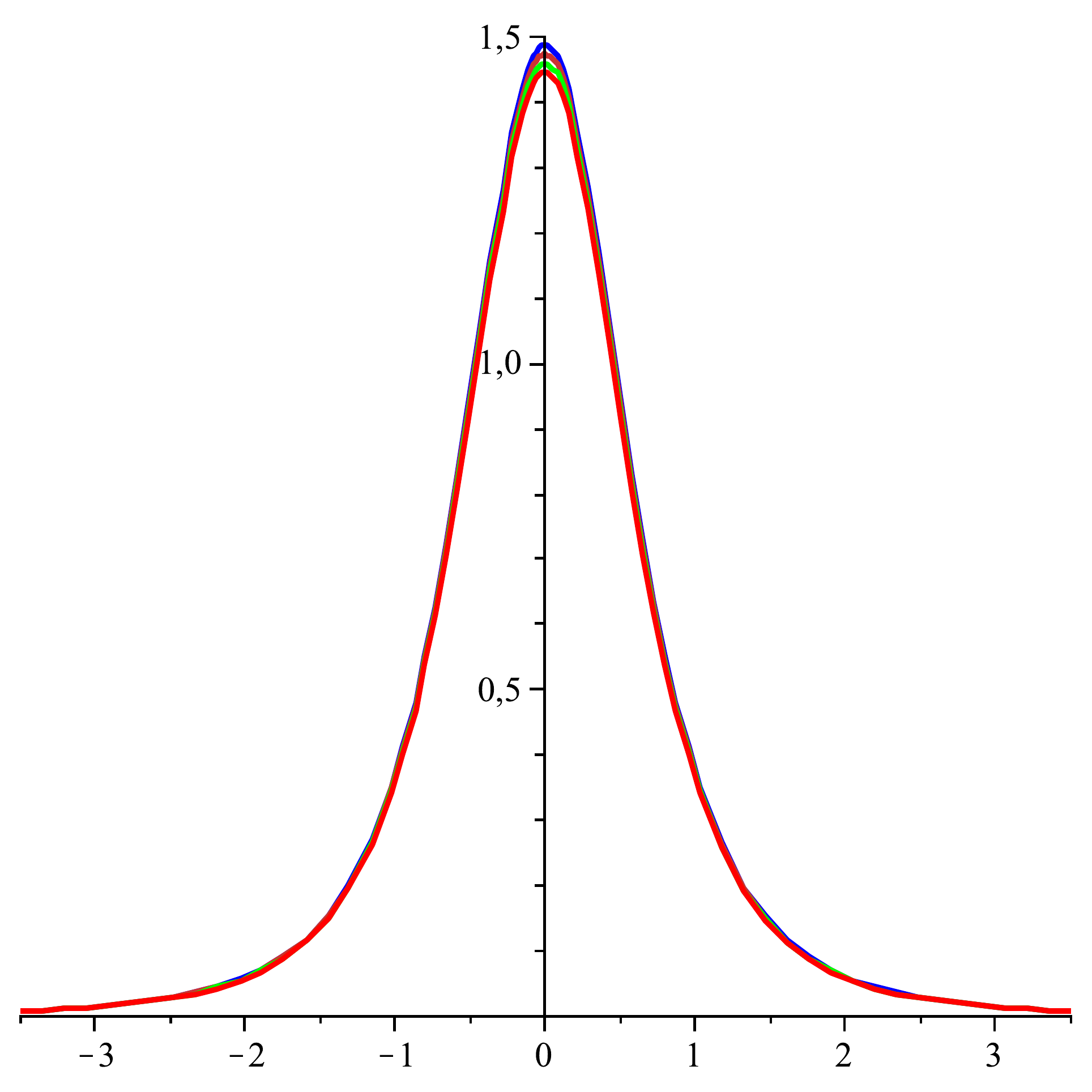}
\caption{\small
    Plots of $B(x)$ (left) and $V(x)$ (right) for asymmetric configurations with
    $m =0$, $\alpha = - \pi/20$,
    $p = 0.1$ and $Q = 1.01,\ 1.02,\ 1.03,\ 1.04$ (bottom-up). The plots of
    $V(x)$ almost merge for these close values of the parameters.}\label{f4}
\bigskip
\end{figure*}
\begin{figure*}
\centering
\includegraphics[width=6cm,height=4.5cm]{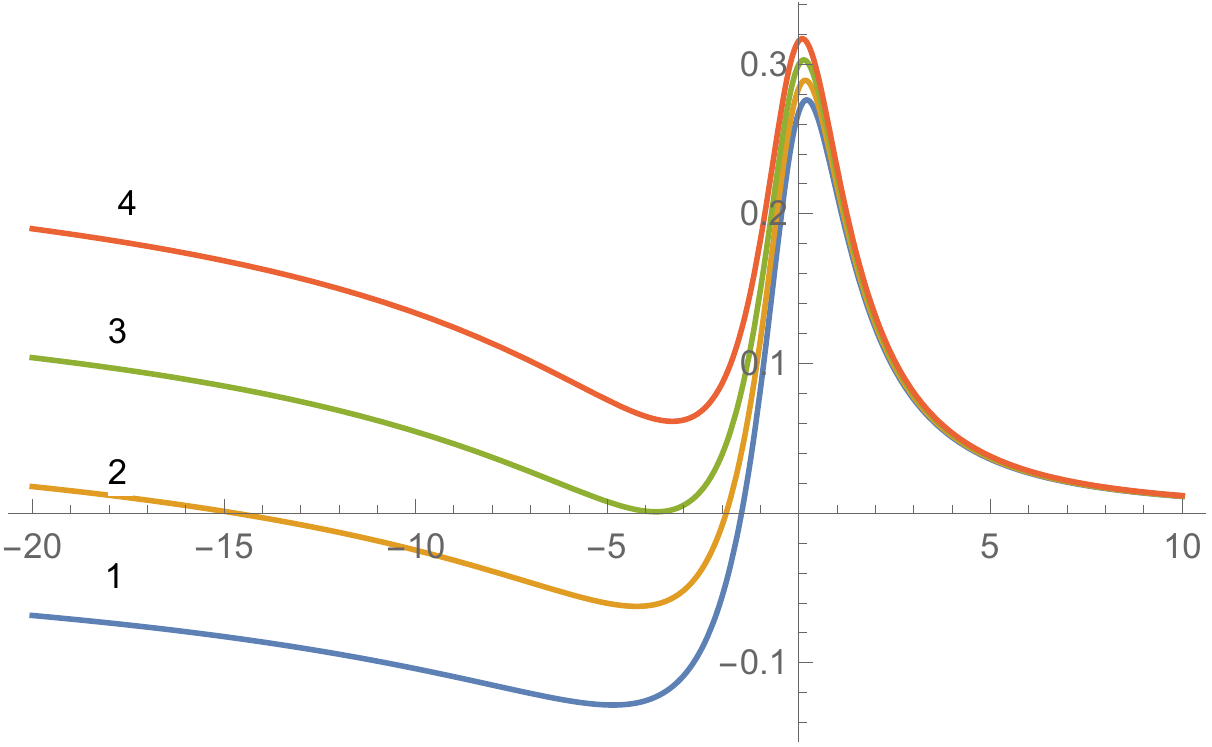}
\
\includegraphics[width=5.5cm,height=4.5cm]{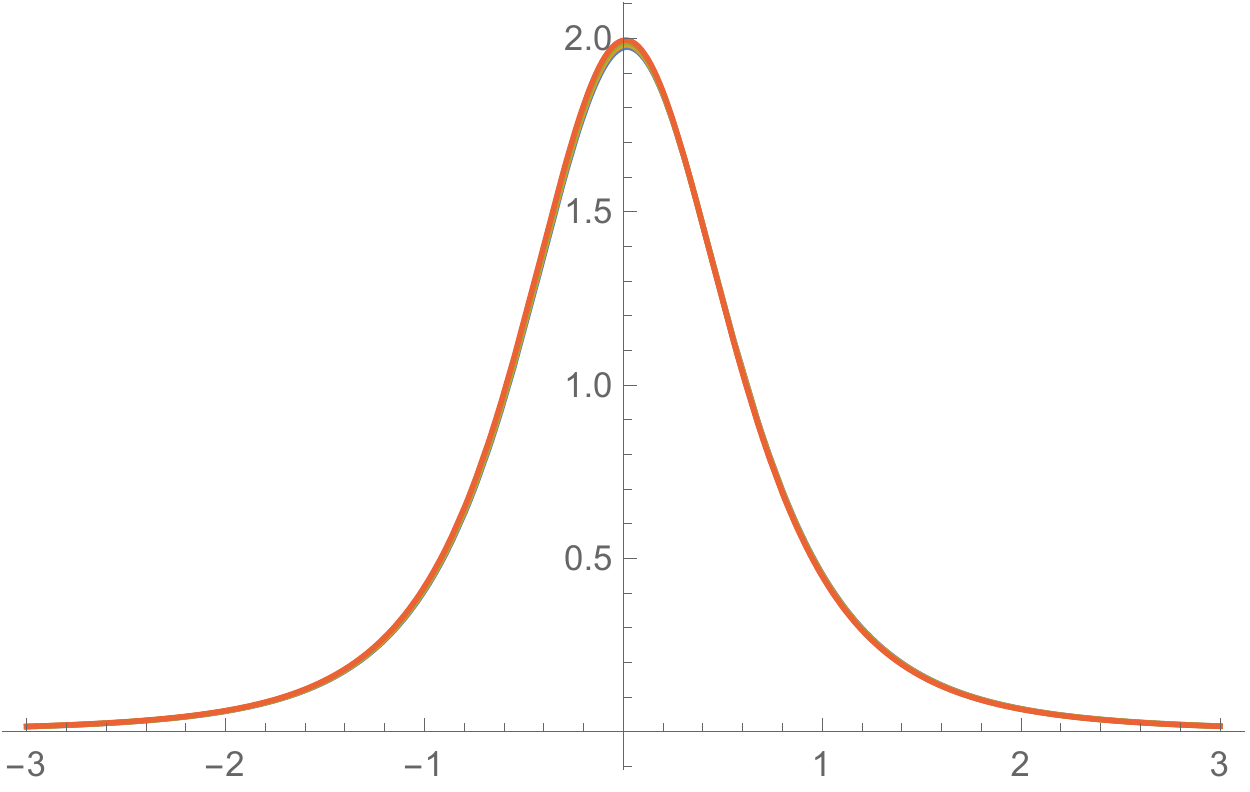}
\
\includegraphics[width=5.5cm,height=4.5cm]{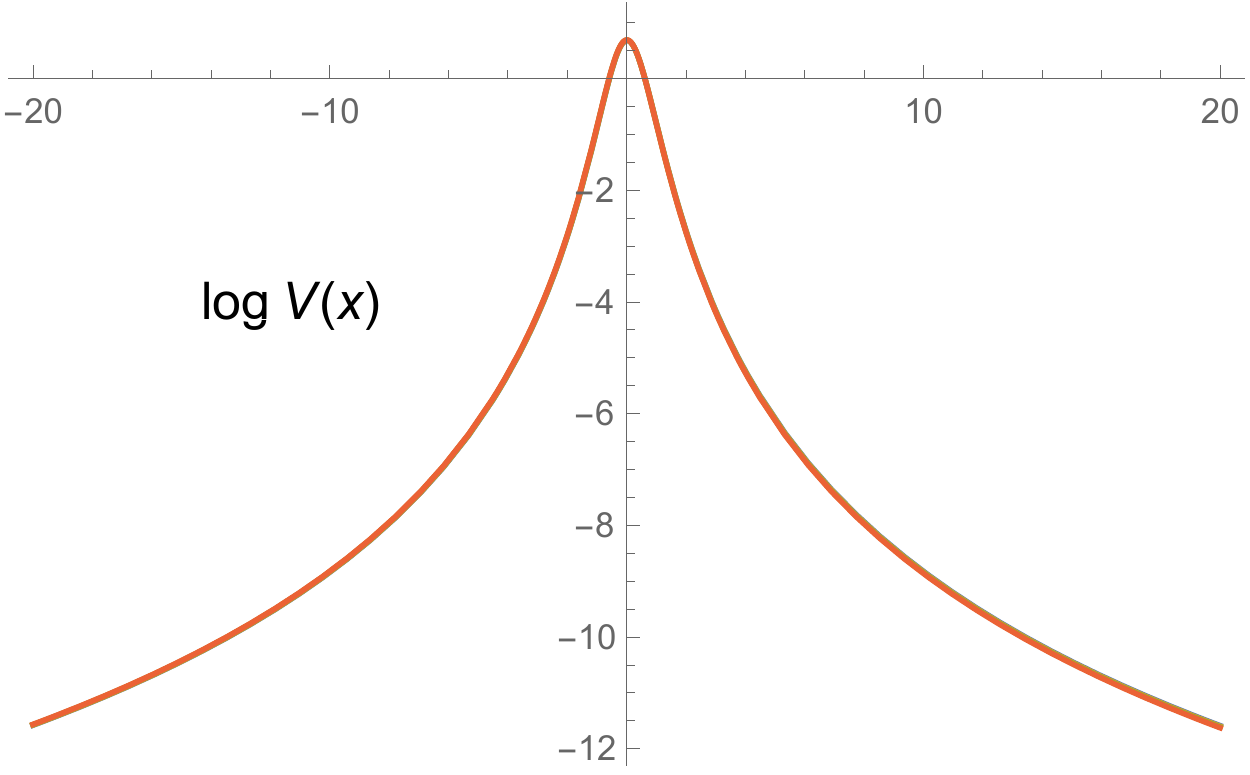}
\caption{\small
    Plots of $B(x)$ (left) and $V(x)$ (middle and right) for asymmetric configurations with
    $\alpha = - \pi/48$, $p = 0.5$ and $Q = 0.813,\ 0.817,\ 0.821,\ 0.825$ (bottom-up). The plots of
    $V(x)$, as in other examples, almost merge. The right panel shows $\log V(x)$ in a wider
    range of $x$, which is possible due to $V >0$.}\label{f5}
\bigskip
\end{figure*}

\subsection{Asymmetric configurations with even $r(x)$}

  The condition $\alpha = 0$ singles out the branch of our general solution with the even function\\
  $r(x) = b\sqrt{x^2 + 1}\cos (p \arctan x)$, in which case we have, in addition to the length scale
  $b$, two free parameters $p$ and $Q$ and the limit \rf{B-} of $B(x)$ at negative infinity.
  Accordingly, we obtain two types of qualitatively different asymmetric configurations: M-AdS
  wormholes in which $B(-\infty) > 0$ (they correspond to larger charges $Q$ at given $p$)
  and M-dS black universes with a single simple horizon and $B(-\infty) < 0$
  (at smaller $Q$ for given $p$). The corresponding functions $B(x)$ and $V(x)$ for
  some values of $p$ and $Q$ are plotted in Fig.\,3. The two kinds of solutions are separated
  by the completely symmetric M-M wormhole solutions discussed in the previous subsection.

\subsection{Asymmetric solutions with $\alpha \neq 0$}

  In this general case there are three parameters that affect the solution behavior, and
  accordingly the properties of geometries, qualitatively determined by the behavior of
  the function $B(x)$, are more diverse.

  From $m\geq 0$ $(a\geq 0)$ it follows $\alpha \geq - \pi p/2$.
  Hence we should study two ranges of $\alpha$:
  (i) $-\pi p/2 \leq \alpha < 0$ and (ii) $0 < \alpha < \pi (1 {-} p)/2$.

  {\bf 1.} At the lower bound,  $\alpha = - \pi p/2$  $(p < 1/2)$, the \Scw\ mass is zero, $m=0$.
  The solution is \asflat\ under the condition $Q > 1$, and the only kind of configurations
  it describes are massless M--AdS wormholes. Examples of the behavior of $B(x)$
  and $V(x)$ in this case are presented in Fig.\,4.

\begin{figure*}
\centering
\includegraphics[width=7.5cm,height=5.5cm]{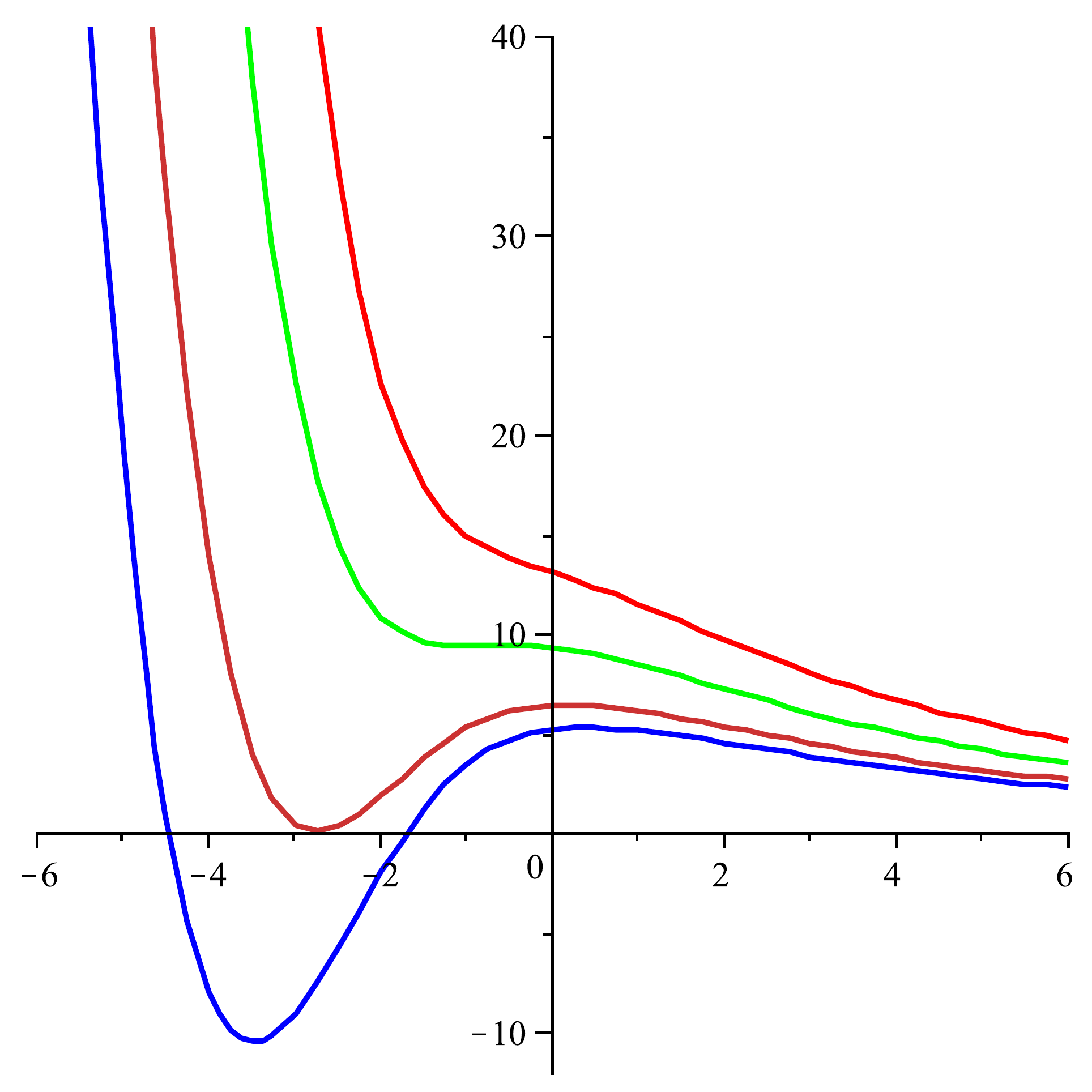}
\qquad
\includegraphics[width=7.5cm,height=5.5cm]{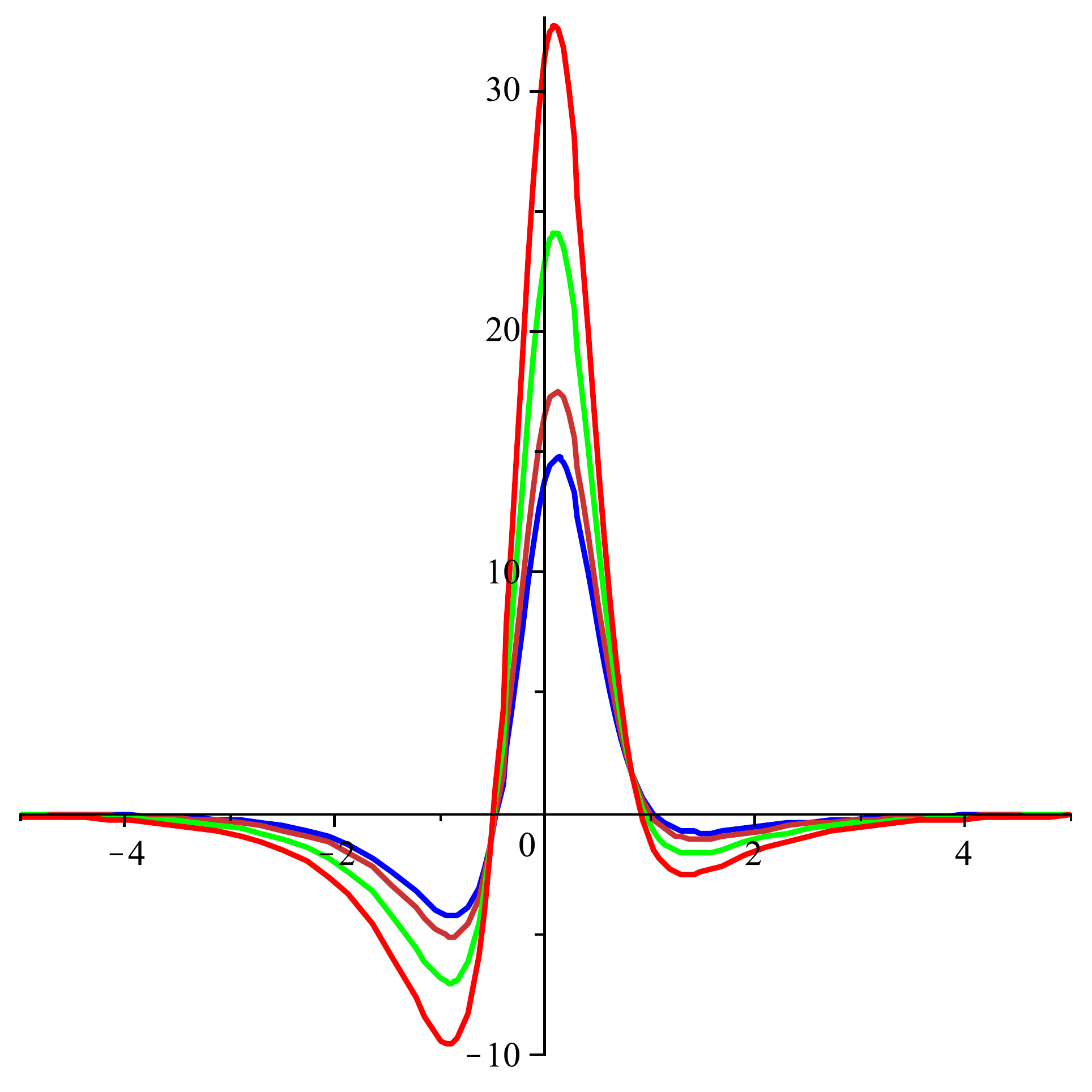}
\caption{\small
    Plots of $B(x)$ (left) and $V(x)$ (right) for singular horn-like configurations with $\alpha = - \pi/20$,
    $p = 0.9$ and $Q = 0.37,\ 0.409,\ 0.5,\ 0.62$ (bottom-up both on the left panel and on the peak
    near $x=0$ on the right panel).}\label{f6}
\bigskip
\end{figure*}

  {\bf 2.} Consider a few examples with $\alpha < 0$ and $m >0$.
  For some such examples we take $\alpha = - \pi/48$ and $p = 0.5$, which implies
  $Q > 0.56$. Trying different values of $Q$, we find its critical values that
  separate different modes of $B(x)$, as illustrated in Fig.\,5.
  At smaller $Q$ (left panel, curve 1) we obtain a black universe (M-dS) with a single horizon.
  Larger values of $Q$ (curve 2) lead to regular M-AdS black holes with two simple horizons
  and a Reissner-Nordstr\"om-like global causal structure (with an AdS asymptotic
  at the far end, $x\to -\infty$, instead of a central singularity). Further slightly increasing
  the $Q$ value, we obtain a regular M-AdS black hole with a double horizon (curve 3).
  And lastly, at larger $Q$ we obtain M-AdS wormholes (curve 4).

  An analysis shows that no other types of the behavior of $B(x)$
  emerge at any admissible values of $\alpha$, both positive and negative,
   leading to finite $B(-\infty)$.

  {\bf 3.} In the limiting case $\alpha = (p - 1)\pi/2$, \eq \rf{25} shows that
  the solution describes a horn-like configuration with  $r \to bp = \const$
  as $x\to - \infty$. In this case, due to the summand proportional to $\sec^{4}z$
  in the expression for $B(x)$, at large negative $x$ we have
  $B (x) \sim  x^4 \to \infty$. Fig.\,6 illustrates the properties of such
  configurations with $\alpha = - \pi/20$ and $p = 9/10$.
  The metric function $A(x)$ also blows up as $x^4$ at the end of the ``horn'', hence
  actually the distance to this end from any point is finite,
   $l = \int dx/\sqrt{A(x)} \sim \int dx/x^2 < \infty$,
  and an inspection shows that there is a curvature singularity. The latter is
  repulsive, and, like the Reissner-Nordstr\"om singularity, can be naked (see
  the upper two curves in Fig.\,6, left) or hidden beyond one extremal or two simple
  horizons (the lower two curves).

\section{Concluding remarks}

  We have found a family of exact static, spherically symmetric solutions in the Einstein-Cartan
  theory (ECT) of gravity, with sources in the form of a nonminimally coupled non-phantom
  scalar field and an electromagnetic field. From this whole family we have selected
  asymptotically flat solutions with a nonnegative \Scw\ mass. The remaining subfamily depends
  on four constants: the nonminimal coupling coefficient $\xi > 1/2$, an arbitrary length scale
  $b > 0$ and two significant integration constants: the dimensionless
  electromagnetic charge $Q$ and the ``asymmetry factor'' $\alpha$.
  With different values of these parameters, the solution describes
  (i) twice \asflat\ (or M-M) symmetric \whs, (ii) asymmetric M-AdS \whs\ with zero or nonzero
  masses, (iii) regular M-AdS black holes with an extremal horizon or two simple horizons, and
  (iv) M-AdS black universes. It is important that, in all these solutions with a nonsingular metric,
  the torsion scalar also remains finite in the whole space, unlike the recently obtained
  \wh\ solutions in the ECT with two scalars \cite{br-gal15}.

  In addition, there are a number of singular solutions that remained beyond the scope of this study,
  for example, those with $\alpha$ outside the range \rf{25a}. It seemed that if $\alpha$ takes
  a marginal value such that  the spherical radius tends to a finite constant $r=r_0$ at the ``far end'',
  $x\to -\infty$. It turns out, however, that such solutions possess a repulsive singularity at $r=r_0$,
  and, depending on $Q$, this singularity can be naked or hidden beyond horizons from the viewpoint
  of a distant observer.

  It is clear that in our solutions the existence of a minimum of $r(x)$ (a throat) is 
  provided by WEC and NEC violation by the effective SET (\ref {10}) of the 
  scalar--torsion field. A question of interest is whether or not the purely scalar SET
  respects the energy conditions, so that their violation might be completely ascribed
  to the contribution of torsion. We notice, however, that this question cannot be asked
  correctly because only the effective SET $T^{i \rm (eff)}_k [\phi]$ satisfies
  the conservation law \rf{11}, and the purely scalar contribution cannot be unambiguously 
  separated from that of torsion. 

  Nevertheless, let us try to calculate the quantity $\rho + p_r \equiv T^t_t - T^x_x$
  (whose negative value indicates NEC violation) for the purely scalar contribution in 
  the SET \rf{7}, excluding there the terms containing $S$ and its derivatives, which 
  would be a correct expression for the scalar field SET in the absence of torsion. 
  We obtain
\beq
     \rho[\phi] + p_{r} [\phi] =
     \frac{A(x)\bigl [1 -2\xi+ (1 + 4\xi )x^2 \bigr]}{\kappa \xi b^2 (x^2  + 1)^2}.
\eeq
  This expression is negative at small $x$ (near the throat) since $\xi > 1/2$, but becomes 
  positive far from it. This resembles the concept of a ``trapped ghost'' \cite{BDon11, BS10}, 
  but, in contrast to these papers, here the kinetic term of the scalar field does not change its sign.  
  
  It is of interest that the effective density  $\rho^{\rm (eff)}[\phi] = T^{t \rm (eff)}_t$ 
  can be positive near the throat in our solution, so that NEC violation is caused by a larger 
  negative value of the effective radial pressure $p_r^{\rm (eff)}[\phi] = -T^{x \rm (eff)}_x$
  Thus, for symmetric configurations $(\alpha = 0,\ K = 0)$ we obtain at $x=0$
\beq
               \rho^{\rm (eff)}[\phi] =  \frac{1}{\kappa b^2} \Bigr[2 + 2(p^2 -1)B_0 + Q\Bigr].
\eeq
  For $p = 0.5$ \, $(a = Q = \pi/4)$ we have
\beq
  \rho^{\rm (eff)}[\phi] =  \frac{1}{\kappa b^2 } \Bigr[6 + 4\ln 2  - \frac{3\pi }{4}\Bigr ] > 0 .
\eeq

  We conclude that the ECT, like some other extensions of GR, provides the existence of regular
  configurations without a center (\whs, \bus, and regular black holes with two asymptotic regions)
  with normal (non-phantom) fields. It means that torsion here replaces (or plays the part of)
  exotic matter while forming such regular objects. A feature of interest in the present solution is
  the necessity of the \elmag\ field for obtaining asymptotic flatness, but it is evidently a property
  of the present family of solutions rather than a general property of the theory.

  A challenging problem is that of stability of these and other solutions of the ECT, and we
  hope to deal with it in our further studies.

\subsection*{Acknowledgments}

  The work of KB was partly performed within the framework of the Center 
  FRPP supported by MEPhI Academic Excellence Project 
  (contract No. 02.a03.\\ 21.0005, 27.08.2013).
  This paper was also financially supported by the Ministry of Education and Science of the Russian
  Federation on the program to improve the competitiveness of the RUDN University among the 
  world leading research and education centers in 2016--2020, and by RFBR grant 16-02-00602.

\end{document}